\documentclass[twocolumn,twocolappendix,usenatbib,useAMS,fleqn,usenames,dvipsnames, tighten]{aastex63} 
\usepackage{amsmath}
\hypersetup{linkcolor=MidnightBlue,citecolor=MidnightBlue,filecolor=MidnightBlue,urlcolor=MidnightBlue}
\graphicspath{{./}{figures/}}
\submitjournal{ApJ}
\usepackage{enumitem,amsmath,amstext,mathtools}

\hypersetup{linkcolor=MidnightBlue,citecolor=MidnightBlue,filecolor=MidnightBlue,urlcolor=MidnightBlue}

\begin{document}

\title{Phase-dependent magnetic coherence in the turbulent interstellar medium}
\shorttitle{Phase-dependent magnetic coherence}
\shortauthors{Butsky et al.}

\correspondingauthor{Iryna S. Butsky}
\email{ibutsky@stanford.edu}

\author[0000-0003-1257-5007]{Iryna S. Butsky}\thanks{NASA Hubble Fellow}
\affiliation{Department of Physics, Stanford University, Stanford, CA 94305, USA}
\affiliation{Kavli Institute for Particle Astrophysics \& Cosmology, P.O. Box 2450, Stanford University, Stanford, CA 94305, USA}

\author[0000-0002-9961-2984]{Caleb Redshaw}
\affiliation{Department of Mechanical Engineering, Stanford University, Stanford, CA 94305, USA}

\author[0000-0002-2679-4609]{Minjie Lei}
\author[0000-0002-7633-3376]{Susan E. Clark}
\affiliation{Department of Physics, Stanford University, Stanford, CA 94305, USA}
\affiliation{Kavli Institute for Particle Astrophysics \& Cosmology, P.O. Box 2450, Stanford University, Stanford, CA 94305, USA}
\author[0000-0003-3806-8548]{Drummond B. Fielding}
\affiliation{Department of Physics, New York University, New York, NY 10003, USA}


\begin{abstract}
Magnetic fields permeate the multiphase interstellar medium (ISM), yet their phase-dependent structure remains poorly constrained by observations. Dust polarization and \ion{H}{1} emission together offer complementary probes of the plane-of-sky magnetic field and cold neutral medium (CNM) gas structure, respectively. Recent observational work has shown that in the diffuse ISM, the dust polarization fraction correlates positively with the CNM mass fraction ($f_{\rm CNM}$) but not with total \ion{H}{1} column density, suggesting a phase-dependent magnetic field geometry. Here, we use extremely high-resolution ($2048^3$) simulations of the turbulent, magnetized, multiphase ISM to investigate the physical origin of this trend. By constructing synthetic \ion{H}{1} and dust polarization maps, we directly compare our simulations to the observational results of \citet{Lei:2024}. We recover a positive $f_{\rm CNM}$--polarization correlation most clearly for sightlines intersecting fewer than $\sim$20 discrete CNM clouds, while the trend becomes weak or intermittent for larger cloud counts, consistent with the expectation that high-Galactic-latitude sightlines contain relatively few independent cold structures. We show that this correlation reflects genuine phase-dependent magnetic structure: CNM clouds tend to be elongated along the local magnetic field and, when normalized by column density, exhibit lower magnetic disorder than the warm neutral medium (WNM). We further demonstrate that apparent discrepancies between simulation- and observation-based measures of magnetic disorder arise from whether disorder is quantified per unit path length or per unit mass. Our results support a picture in which CNM structures host relatively ordered magnetic fields, producing higher polarization fractions along CNM-dominated sightlines in the diffuse ISM.
\end{abstract}

\section{Introduction}
Magnetic fields are ubiquitous in the Milky Way, other galaxies, and the intergalactic medium, playing pivotal roles across various cosmic scales. In the interstellar medium (ISM), magnetic fields help govern gas dynamics and direct the trajectory of cosmic rays \citep[e.g.,][]{Crutcher:2012, Zweibel:2017}. Furthermore, the contaminating foreground of the ISM magnetic field directly interferes with the detection of primordial B-mode polarization of the cosmic microwave background (CMB), which would give the first direct evidence of inflation \citep{Kaminkowski:1997, Seljak:1997a, Seljak:1997b}. Thus, understanding the Milky Way magnetic field structure is crucial for deciphering the complex dynamics of the ISM, understanding magnetism in other galaxies, and improving our understanding of cosmology. Despite its importance, determining the three-dimensional structure of the Galactic magnetic field and its phase-dependent variations within the ISM is challenging due to the line-of-sight (LOS) integrated nature of most astrophysical tracers \citep[e.g.,][]{Ferriere:2001}. 

Although we cannot directly observe magnetic fields, recent advancements have significantly enhanced our ability to infer their structure using indirect tracers. For example,
observations of the neutral ISM through \ion{H}{1} 21 cm
emission provide three-dimensional (position-position-velocity) insights into the cold neutral
medium (CNM), warm neutral medium (WNM), and
thermally unstable medium (UNM), though decomposing \ion{H}{1} emission into contributions from these thermal phases is a notoriously difficult problem. High-resolution \ion{H}{1} emission maps show that the diffuse neutral ISM is dominated by thin filamentary features within the CNM that are aligned with the magnetic field \citep{McClureGriffiths:2006, Clark:2014, Clark:2015, Kalberla:2016, Peek:2019}. Additionally, polarized thermal emission from dust grains traces the plane-of-sky (POS) component of the interstellar magnetic field, integrated over the LOS. 
Combining \ion{H}{1} emission and dust polarization can provide insights into the relative tangling of the magnetic field within different phases of the neutral ISM \citep[e.g.,][]{Clark:2018}.

However, interpreting the relative disorder of the magnetic field in different ISM phases is far from straightforward. Using maps of the CNM mass fraction \citep{Murray:2020} estimated from GALFA-\ion{H}{1} emission \citep{Peek:2018} and Planck dust polarization data \citep{Planck:2020a}, \citet{Lei:2024} find a strong correlation between the dust polarization fraction and CNM mass fraction in the diffuse ISM, where there is no correlation between the polarization fraction and total \ion{H}{1} column density. The authors show that these data are consistent with a phase-dependent magnetic field interpretation, where the magnetic field is more ordered in the CNM and UNM per unit column density, and relatively disordered in the WNM. On the surface, these predictions seem to contradict the interpretations presented in \citet{Ghosh:2017} and \citet{Adak:2020}, which interpret the magnetic field in the CNM to be more turbulent than in the WNM. Although it is possible that the apparent discrepancy is due to the differences in methodology or regions of sky analyzed, understanding the true physical structure and phase distribution of the ISM magnetic field necessitates comparison to state-of-the-art simulations.

Simulating this phase dependence of magnetic fields in the ISM presents significant challenges due to the vast range of scales involved. Faithfully comparing simulations to complex observations not only requires the inclusion of magnetism but also necessitates incorporating models of radiative cooling and heating. Moreover, simulations must achieve sufficiently high resolution to capture the multiphase nature of the ISM. Until recently, meeting these requirements was computationally prohibitive. In this study, we utilize the extremely high-resolution simulations of the turbulent, magnetized ISM detailed in \citet{Fielding:2023}, from which we construct synthetic observables that enable direct comparison with observations. This approach allows us to draw connections between observed polarization and CNM column density correlations and the physical three-dimensional structure of the magnetic field. 

The structure of this paper is as follows. \autoref{sec:methods} describes the simulations and the construction of synthetic observations. \autoref{sec:results} presents the main results, \autoref{sec:discussion} discusses their implications, and \autoref{sec:summary} summarizes our conclusions.

\begin{figure}
    \includegraphics[width = 0.48\textwidth]{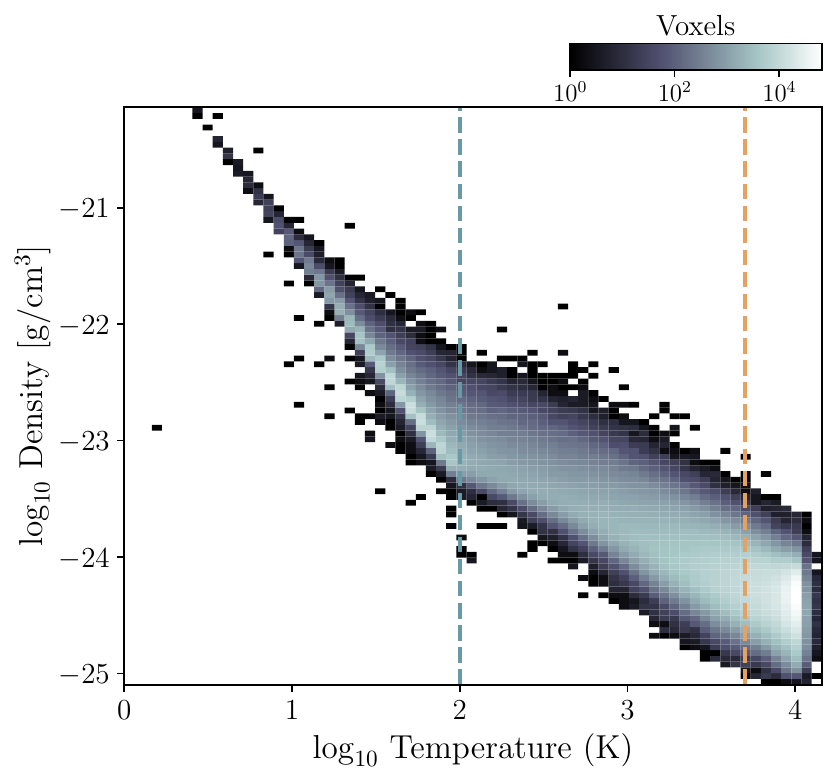}
    \caption{Two-dimensional density plot of density versus temperature for each voxel in the randomly selected sightlines through the simulation volume. Dashed lines indicate fiducial phase thresholds for CNM ($T < 100 \mathrm{K}$; blue) and WNM ($T > 5000 \mathrm{K}$; orange).}
    \label{fig:Tdensphaseplot}
\end{figure}

\begin{figure*}
\includegraphics[width=0.9\textwidth]{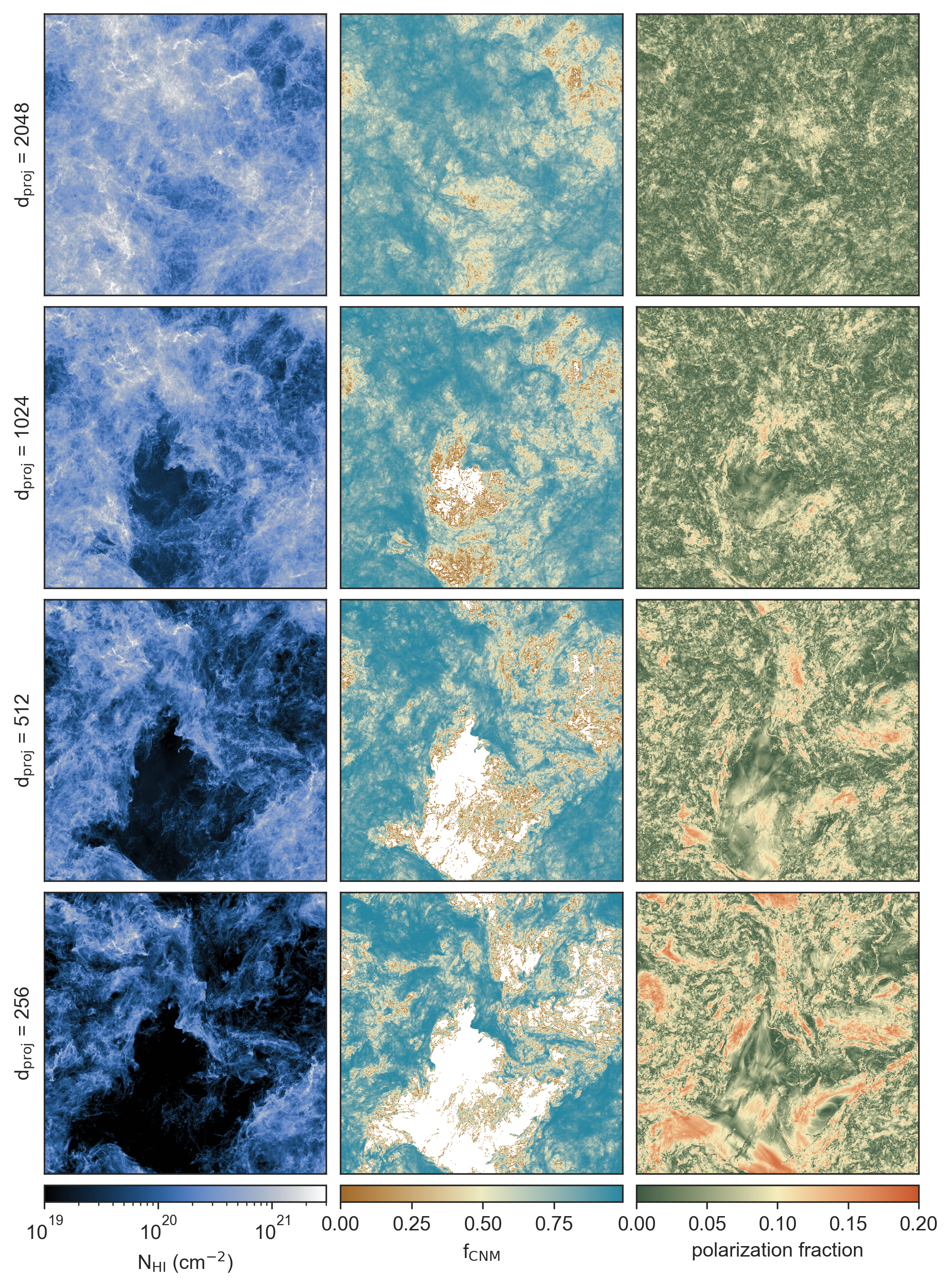}
\caption{Total \ion{H}{1} column density, cold mass fraction ($f_{\rm CNM}$), and polarization fraction plotted against varying projection depths from the full simulation box depth to reduced depths. Regions colored in white highlight where the CNM mass fraction is zero. The trend shows a decrease in $f_{\rm CNM}$ and an increase in polarization fraction with reduced depth, reflecting observational characteristics of the high Galactic latitude sky with fewer cold gas structures along sightlines. }
\label{fig:HIcol}
\end{figure*}

\begin{figure*}
    \includegraphics[width = \textwidth]{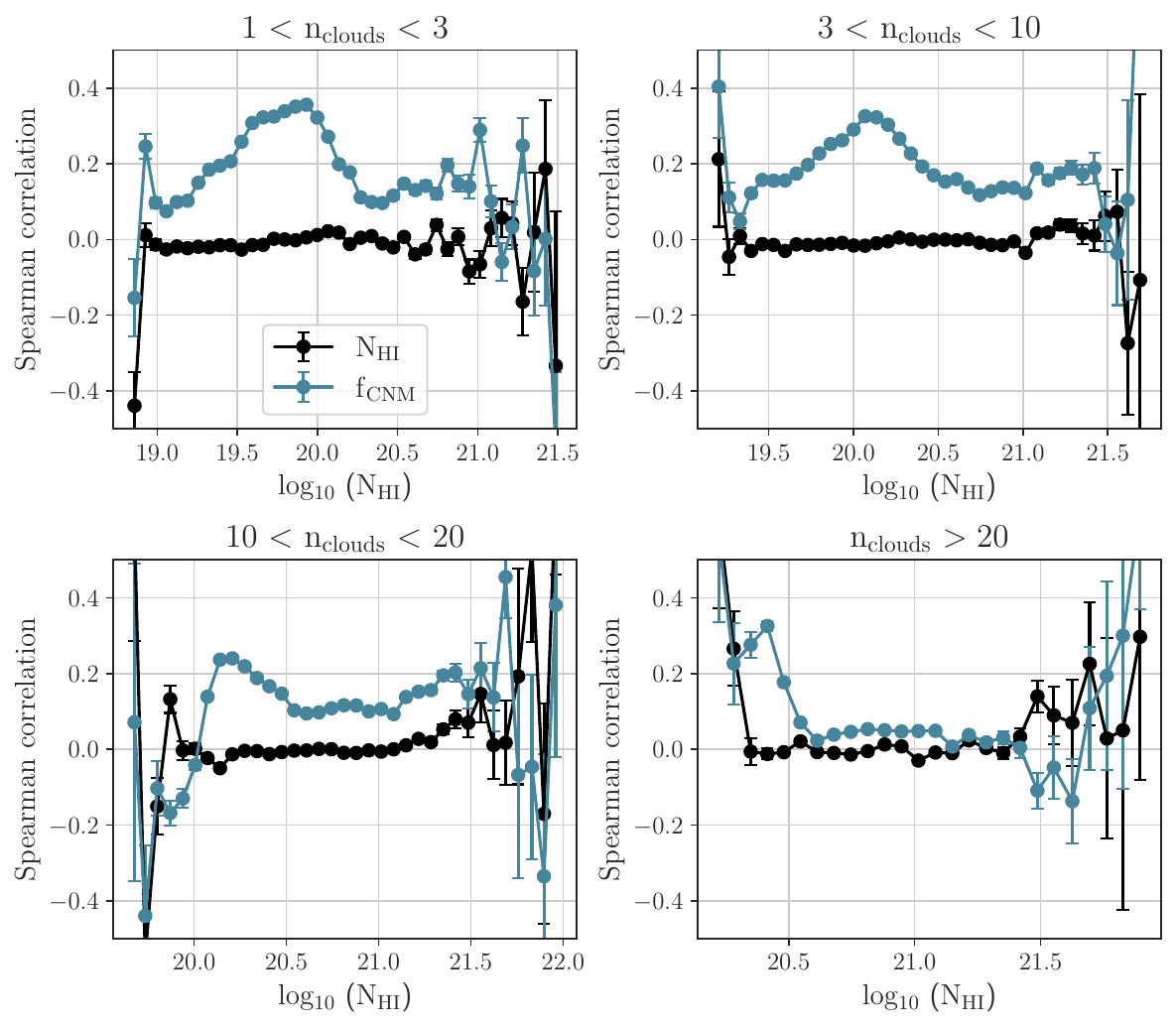}
    \caption{Correlation between dust polarization fraction and total \ion{H}{1} column density (black) and $f_{\rm CNM}$ (blue) as a function of binned $N_{\rm HI}$. To probe different effective CNM depths, we include projections spanning the range of path lengths shown in \autoref{fig:HIcol}. Each panel corresponds to sightlines intersecting a different number of cold clouds. Error bars indicate the standard deviation of the Spearman correlation coefficient across 100 bootstrap resamplings within each $N_{\rm HI}$ bin. The $f_{\rm CNM}$--polarization correlation is strongest for sightlines intersecting fewer than $\sim$20 cold clouds; for $n_{\rm clouds}\gtrsim 20$ it becomes weak or intermittent, appearing only over limited $N_{\rm HI}$ intervals and disappearing across much of the sampled column-density range.}
    \label{fig:correlation}
\end{figure*}

\section{Numerical Methods}\label{sec:methods}
\subsection{Simulations}

We use the high-resolution turbulent box simulations presented in \citet{Fielding:2023}, as well as synthetic observations forward-modeled from those simulation data. Here we briefly summarize the simulation properties. The turbulence is driven on the largest scales with a root-mean-square sonic Mach number $M^2 = \left<v^2\right>/\left<c_s^2\right> = 1$ and equal power injected into wavenumbers $kL/2\pi = 1$ and 2. Similarly, the tangled magnetic fields are initialized with a root-mean-square plasma beta $\beta = \left<P\right>/\left<B^2/2\right> = 1$ and with perturbations at large scales with wavenumbers $kL/2\pi = 1$ and 2. 
The simulations were run with the \texttt{athena++} code \citep{athena}. We focus our analysis on the highest-resolution box, resolved with $2048^3$ cells.

The simulation domain is initialized with a constant thermal pressure ($P_0$), gas density ($\rho_0$), and unstable equilibrium temperature ($T_0$). The simulations include idealized radiative cooling and heating terms that mimic the cooling and heating rates of the Milky Way ISM. The radiative cooling is estimated with a simple piecewise power-law function of temperature (i.e., $\Lambda \propto T^\alpha$ in each segment) to approximate the shape of the cooling curve, while the heating rate $\Gamma$ is constant, normalized such that the gas is unstable at the initial conditions $P_0, T_0$ and forms two stable equilibrium temperatures, $T_{\rm cold} = T_0/100$ and $T_{\rm warm} = 10T_0$. 

While the simulations are formally scale-free, our choice of physical cooling time in post-processing sets the physical size of the domain to $L\approx 75$ pc. We analyze the simulation at $t = 22\,\rm{Myr}$, which corresponds to 2.24 eddy turnover times in the warm gas. By this time, the system displays properties consistent with three thermal phases—the warm, unstable, and cold neutral media—with tangled magnetic fields permeating the volume. We note that the adopted cooling curve does not explicitly track ionization, so the warm phase should be interpreted as a generic $\sim 10^4$ K component rather than a distinct separation between warm neutral and warm ionized media. \autoref{fig:Tdensphaseplot} shows the volume distribution of a random sample of voxels in density–temperature space.

We refer the reader to \citet{Fielding:2023} for additional details on the simulations. 

\subsection{Analysis}
\subsubsection{Synthetic \ion{H}{1} Maps}

We compute \ion{H}{1} column densities by integrating the neutral hydrogen density along the simulation LOS. For simplicity, we align our analysis with the $z$-axis of the simulation domain, such that
\begin{equation}\label{eq:NHI}
N_{\rm HI} = \sum \rho_{\rm HI} {\rm d}z,
\end{equation}
where $\rho_{\rm HI}$ is the mass density of \ion{H}{1}. 

We estimate the dust polarization fraction as 
\begin{equation}
p = \frac{\sqrt{Q^2+U^2}}{I},
\end{equation}
where
\begin{equation}
Q = \int\alpha\epsilon\rho\cos2\psi\cos^2\gamma ds,
\end{equation}
\begin{equation}
U = \int\alpha\epsilon\rho\sin2\psi\cos^2\gamma ds,
\end{equation}
and
\begin{equation}\label{eq:StokesI}
I = \int\epsilon\rho ds - 1/2\int\alpha\epsilon\rho(\cos^2\gamma-\frac{2}{3}) ds.
\end{equation}
Here, $\epsilon=1$ is the emissivity; $\rho$ is the local gas density; $\psi$ is the magnetic field angle in the POS (here, the x-y plane); $\gamma$ is the inclination angle between the magnetic field and the POS; and the grain alignment parameter $\alpha$ is determined by $\frac{\alpha}{1-\alpha/6}=p_{max}=0.2$ \citep{Fiege:2000, Pelkonen:2007}.

We also construct single-phase versions of the column density and polarization maps. In this case, we repeat the calculations in Equations~\ref{eq:NHI}–\ref{eq:StokesI}, restricting the sums and integrals to voxels belonging to a given thermal phase. For example, we define the “CNM dust polarization fraction” as the polarization fraction computed using only CNM gas within the simulation volume.

\subsection{Phase-separated ``clouds"}
A typical sightline through the simulation volume is multi-phase, i.e., the LOS intersects gas in each of the cold, warm, and unstable neutral phases. For some of our analysis, we wish to examine the structure of contiguous segments of a sightline that belong to a single phase. We randomly select 1280 sightlines through the simulation volume 
and split each into contiguous one-dimensional segments, or ``clouds", of each phase. Segments that intersect the boundaries of the simulation domain are discarded. We fiducially define as CNM voxels with a temperature $T < 100 \mathrm{K}$, and WNM as $T > 5000 \mathrm{K}$ (Figure \ref{fig:Tdensphaseplot}). Voxels with $100 < T < 5000 \mathrm{K}$ are defined to be in the ``unstable" phase. With these fiducial temperature thresholds, our 1280 random sightlines are partitioned into 22900 cold clouds, 37105 unstable medium clouds, and 60348 warm clouds. Perturbing the temperature thresholds changes the number of clouds assigned to each phase, but does not affect the conclusions in this paper. We comment on the interpretation of these ``clouds" with respect to the observational definition in Section \ref{sec:discussion}.

\begin{figure*}
\centering
\includegraphics[width=\textwidth]{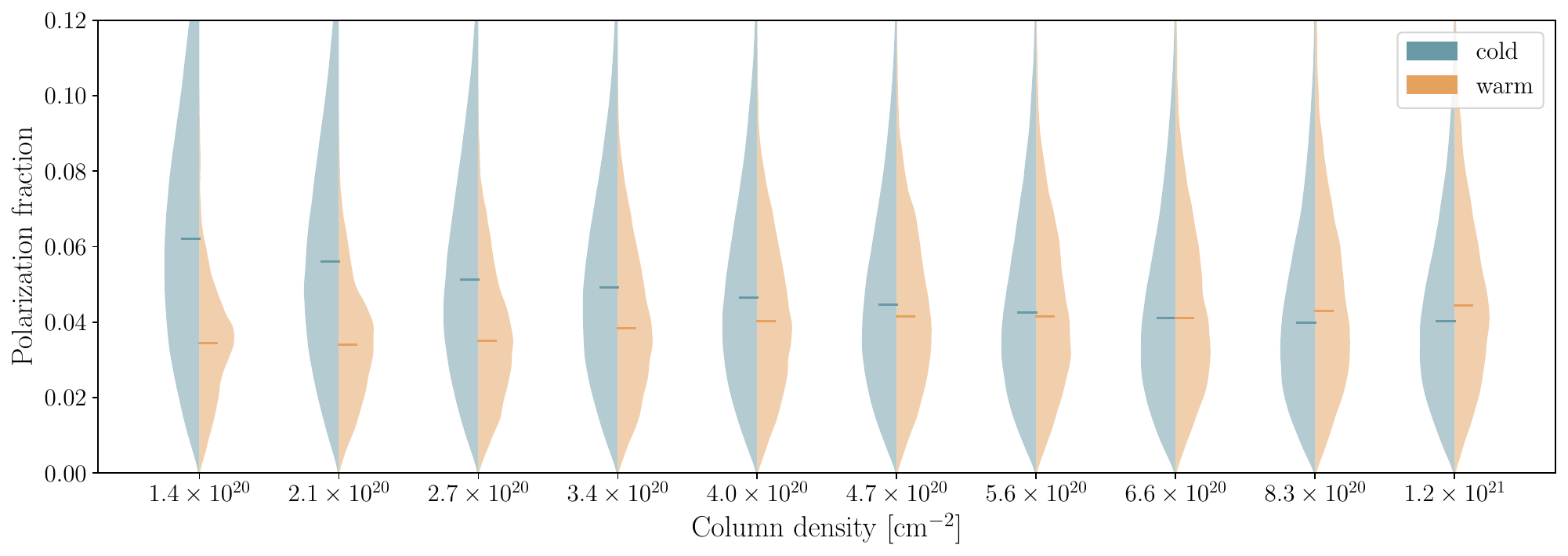}
\caption{Density plots of the dust polarization fraction computed only from voxels belonging to either the cold (teal) or warm (orange) phases. The distributions are plotted in deciles of the total column density along the LOS through the simulation volume, labeled by the median column density of the bin. Horizontal lines indicate the median of each distribution. }
\label{fig:violinplot}
\end{figure*}

\section{Results}\label{sec:results}
Our primary goal in this work is to investigate the physical association between gas in different thermal phases and the structure of the magnetic field. 
We wish to understand how the physical properties of the medium relate to observational signatures in \ion{H}{1} column density and dust polarization fraction. 

\subsection{Observable correlations between CNM mass fraction and dust polarization}

In \autoref{fig:HIcol}, we present the total \ion{H}{1} column density alongside the cold neutral medium mass fraction ($f_{\rm CNM}$) and polarization fraction, varying the projection depth. The initial row utilizes the complete depth of 2048 cells, with subsequent rows representing reduced depths of 1024, 512, and 256 cells, respectively. The shallower depths approximate the observational characteristics of the high Galactic latitude sky, typically featuring fewer cold gas structures per sightline and lower $f_{\rm CNM}$ values \citep[e.g.,][]{McClure-Griffiths:2023}.

At full box depth, we find regions with high $N_{\rm HI}$ and near-unity $f_{\rm CNM}$ values accompanied by low polarization fractions. As the projection depth decreases, regions with reduced $f_{\rm CNM}$ become more prominent, and the polarization fraction approaches the intrinsic maximum value of 0.2. This highlights the importance of selecting appropriate path lengths when comparing turbulent-box simulations to the observed ranges of $N_{\rm HI}$, $f_{\rm CNM}$, and polarization at high Galactic latitudes. 

This approach carries two caveats. First, real observations probe extended sightlines that may include substantial WNM outside CNM-dominated regions; the finite LOS depth of our simulations may therefore under-sample this component and bias the phase balance relative to true Galactic columns. Second, our simulations model a driven turbulent box representative of dynamically active ISM environments and may not fully capture the conditions characteristic of the high–Galactic-latitude sky. Still, we expect that the underlying connection between magnetic field geometry, thermal phase structure, and polarized emission may be robust to other aspects of the physical environment. The advantages and limitations of this simulation for these purposes are discussed further in Section \ref{sec:discussion}.

\autoref{fig:correlation} examines how the correlation between CNM mass fraction and polarization fraction varies with total \ion{H}{1} column density, $N_{\rm HI}$, and with the number of distinct CNM clouds intersected along each sightline. The y-axis shows the Spearman correlation coefficient between the CNM mass fraction ($f_{\rm CNM}$) and the dust polarization fraction. For nearly all sightlines in the range $19 < \log(N_{\rm HI}[\mathrm{cm}^{-2}]) < 20.5$, we find a positive correlation between $f_{\rm CNM}$ and polarization fraction. This trend persists—though over a narrower $N_{\rm HI}$ interval—even for sightlines intersecting more than 20 CNM clouds. The polarization fraction is generally uncorrelated with $N_{\rm HI}$ within the same $N_{\rm HI}$ bins: thus, the correlation between $f_{\rm CNM}$ and the polarization fraction cannot be trivially explained as a correlation between $f_{\rm CNM}$ and $N_{\rm HI}$.

The qualitative behavior of the $f_{\rm CNM}$–polarization correlation is consistent with the observational result reported by \citet{Lei:2024}. We caution, however, that the specific $N_{\rm HI}$ range over which this trend appears in the simulations should not be interpreted too literally, as it depends on the details of the simulation setup (e.g., box size and phase balance) and is not expected to match Galactic sightlines quantitatively.

We can use the existence of this correlation in the synthetic observations to probe its origin in the 3D distribution of magnetic fields and gas phases within the simulation. The simplest implication of this positive correlation is that, per unit column density, the CNM volume contributes more strongly polarized emission than the WNM. Because there is no phase-dependent emissivity in our simulations, this must reflect differences in magnetic field tangling between phases.

\autoref{fig:violinplot} further supports the conclusion that the CNM contributes more strongly to the polarized emission per unit column density than the WNM. The figure shows the polarization fraction computed separately from the CNM (teal) and WNM (orange) gas phases, for sightlines with at least 16 voxels belonging to each phase. The distributions of these phase-restricted polarization fractions, shown as a function of the total column density, indicate that for column densities $\lesssim 6 \times 10^{20}~\mathrm{cm}^{-2}$, the CNM volume exhibits less geometric depolarization than the WNM volume. This plot is qualitatively unchanged if the phase-restricted polarization fractions are analyzed as a function of the number of cold clouds intersected along each sightline. In that case, we find that the polarization fraction associated with the CNM is shifted higher than that of the warm gas for sightlines intersecting fewer than $\sim 25$ clouds. The relative behavior of the cold and warm polarization fraction distributions is similar when we restrict the box projection depth to 1024, 512, or 256 voxels, despite the polarization fraction distribution shifting toward higher values overall for those shallower depths.

\begin{figure}
    \includegraphics[width = 0.48\textwidth]{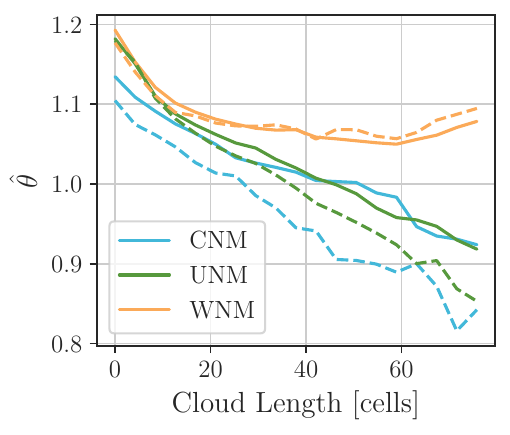}
    \caption{The average angle between the magnetic field direction in a cloud and the LOS as a function of the LOS cloud length. Here, $\hat{\theta} = 0$ would indicate a magnetic field that is perfectly aligned with the LOS and $\hat{\theta} = \pi / 2$ indicates a magnetic field that is perfectly in the POS direction. As expected, the value of $\hat{\theta}$ decreases with increasing CNM cloud length, indicating that the magnetic field is more likely to be aligned with CNM filaments than with the WNM or UNM. The dashed lines show the same measurement for a total \ion{H}{1} column density cut of $19.5 < \log(N_{\rm HI}[\mathrm{cm}^{-2}]) < 20.5$, corresponding to the regions in \autoref{fig:correlation} where $f_{\rm CNM}$ is correlated with polarization fraction. }
    \label{fig:alignment}
\end{figure}

\subsection{Phase dependence of the three-dimensional magnetic field structure}

Next, we aim to understand whether the correlation between $f_{\rm CNM}$ and the polarization fraction implies something about the relative (dis)order of the magnetic field across neutral ISM phases. In \autoref{fig:alignment} we look at the relationship between magnetic field alignment and cloud length for individual CNM, UNM, and WNM clouds along the LOS. 

For computational simplicity, we take $\hat{z}$ to define the LOS direction and compute the length of each distinct cloud along this axis as well as the average angle between the magnetic field direction in the cloud and $\hat{z}$, $\hat{\theta} = \cos^{-1}(\hat{b}_z)$. In our definition, $\hat{\theta} = 0$ indicates a magnetic field that is perfectly aligned with the LOS while $\hat{\theta} = \pi/2$ indicates a magnetic field lying entirely in the POS.

\autoref{fig:alignment} shows that, for CNM clouds, greater LOS cloud lengths correspond to mean magnetic field orientations that are more closely aligned with the LOS. This trend is expected if CNM structures are preferentially elongated along the local mean magnetic field relative to the other phases. In that case, filamentary CNM clouds will appear longest when viewed along their major axis; if this axis is aligned with the magnetic field, such sightlines will also correspond to magnetic fields that are more parallel to the LOS.

\begin{figure*}
\centering
\includegraphics[width=\textwidth]{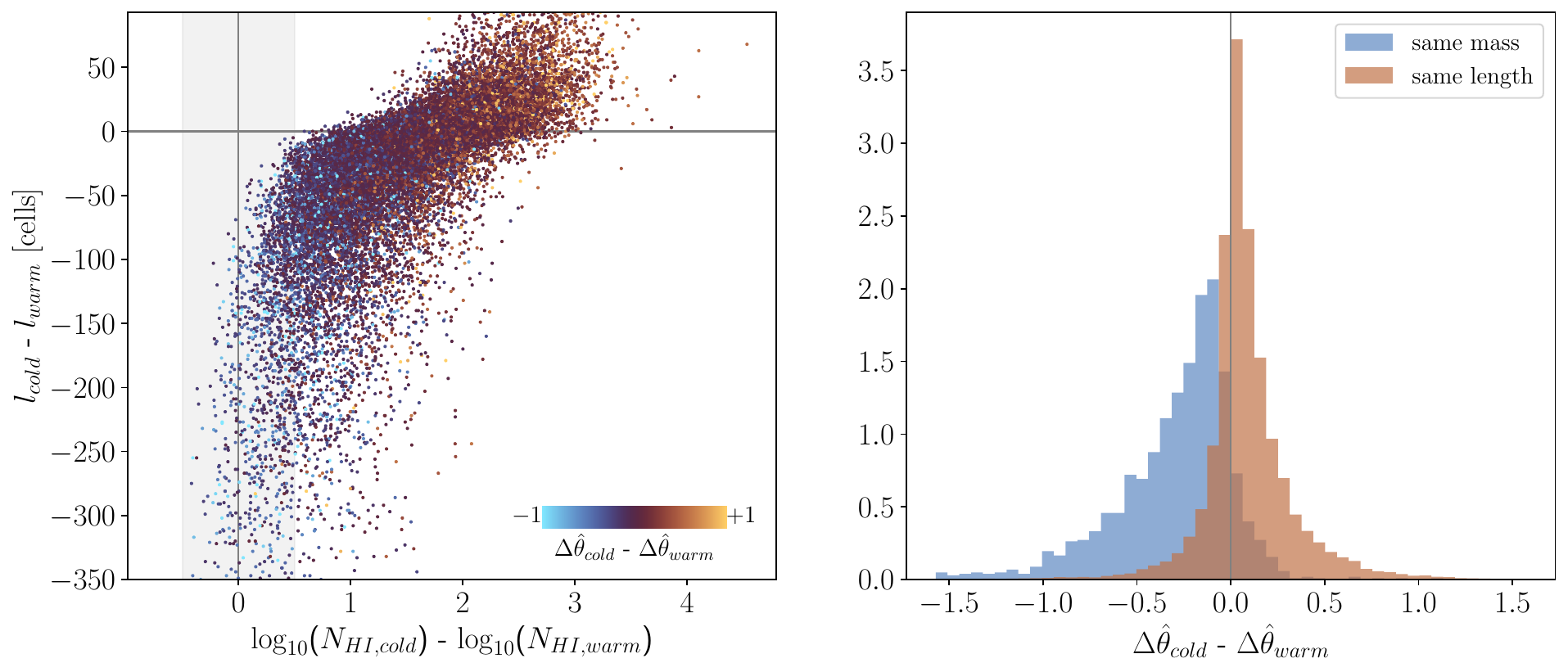}
\caption{Properties of $10^5$ randomly sampled cold and warm cloud pairs. \textit{Left:} The distribution of length difference (cold length minus warm length) vs. mass difference ($\log_{10}$ cold mass minus $\log_{10}$ warm mass). Points are colored by the difference in $\Delta\hat{\theta}$ between the cold and warm clouds (cold minus warm), such that points are more blue (orange) when the cold (warm) cloud has a more ordered magnetic field. For the bulk of the random cold-warm cloud pairs, the cold cloud is more massive, and the warm cloud is longer. \textit{Right:} Normalized histograms of the cold cloud--warm cloud $\Delta\hat{\theta}$ difference for clouds of approximately equal mass (blue) and equal length (orange). The ``same mass" histogram shows randomly sampled pairs within the gray vertical band in the lefthand plot. The ``same length" histogram compares each randomly sampled cold cloud with a same-length \textit{segment} of a random warm cloud. For two clouds of similar masses, the cold cloud is generally shorter and more magnetically ordered. A typical cold cloud is more magnetically disordered than an equivalent-length random segment of a warm cloud.}
\label{fig:lengthmasscompare}
\end{figure*}

Next, we explore the relative disorder of the magnetic field in different gas phases. The left panel of \autoref{fig:lengthmasscompare} compares the difference in the mean magnetic field angle relative to the LOS, $\hat{\theta}$, between randomly paired cold and warm clouds. Each point represents a cold–warm cloud pair and is positioned according to the difference in cloud length and mass, while the color encodes the difference in $\Delta \hat{\theta}$ (cold minus warm), where $\Delta \hat{\theta}$ is defined as the interquartile range (25th to 75th percentile range) of $\hat{\theta}$ within a cloud. We set a minimum cloud length of 8 voxels to be included in this analysis, but note that our conclusions are robust to different choices of this threshold.

For the majority of pairs, the cold cloud is more massive but shorter along the LOS, while the warm cloud is more extended. This reflects the intrinsic structural difference between phases: CNM clouds are denser and occupy smaller characteristic scales, whereas WNM structures are more diffuse and spatially extended. The color distribution indicates that, at fixed mass, the CNM typically exhibits smaller $\Delta \hat{\theta}$ than its warm counterpart, implying greater magnetic coherence per unit mass.

The right panel isolates the roles of mass and spatial scale. The blue histogram compares clouds of approximately equal mass (corresponding to the vertical band in the left panel). In this comparison, cold clouds are generally shorter yet more magnetically ordered. In contrast, the orange histogram compares each cold cloud to a same-length segment drawn at random from a warm cloud. When controlling for LOS extent rather than mass, the CNM often shows larger $\Delta \hat{\theta}$, indicating greater angular variation. We perturb this analysis in a number of ways, and find consistent results. The conclusion that, at fixed length scale, the warm gas is more magnetically ordered than the cold gas is robust when we compare similar-length clouds, rather than subsampling the warm clouds. This conclusion is also unchanged if we set our definition of ``cold gas" to $T < 300~\mathrm{K}$ and ``warm gas" to $T > 8000~\mathrm{K}$. Our interpretation is also robust to the specific choice of statistic used to quantify the magnetic disorder: for instance, we can compare the resultant vector of the cloud magnetic field, normalized to the cloud length, and our conclusions are unchanged.

Taken together, these results suggest that the CNM tends to form shorter, more massive structures that maintain relatively coherent magnetic geometries. The WNM, by contrast, exhibits greater disorder when compared at fixed mass but can appear more ordered when averaged over shorter path lengths. This arises from the different characteristic scales of the two phases. 

This distinction highlights an important difference between simulation-based and observational measures of magnetic disorder. In simulations, where the three-dimensional structure is resolved, it is natural to quantify angular dispersion per unit path length. Observations, however, probe integrated quantities and are therefore sensitive to dispersion as a function of column density—that is, total accumulated mass along the sightline. When magnetic disorder is normalized by mass rather than spatial extent, the CNM contributes less to relative angular variation despite its smaller characteristic scale. Apparent discrepancies between phase-dependent ordering in simulations and observations can thus arise from the metric used to quantify disorder, rather than from a fundamental difference in the underlying magnetic geometry.

\section{Discussion}\label{sec:discussion}
The central question we set out to address is whether our simulations reproduce the observational trends reported by \citet{Lei:2024}—specifically, the positive correlation between the CNM mass fraction ($f_{\rm CNM}$) and the dust polarization fraction in the diffuse sky ($N_{HI} \lesssim 4 \times 10^{20}~\mathrm{cm}^{-2})$. The short answer is yes, with important caveats. 

In our simulations, $f_{\rm CNM}$ and polarization fraction are positively correlated for sightlines intersecting fewer than $\sim$20 discrete CNM clouds. Observationally, this condition is naturally satisfied at high Galactic latitudes, where typical sightlines traverse fewer independent cold structures. By contrast, much of our simulated domain represents a highly turbulent ISM patch with tangled magnetic fields and abundant CNM substructure, more characteristic of inner-disk conditions. \citet{Lei:2024} find that at \textit{higher} column densities ($N_{HI} >10^{21}~\mathrm{cm}^{-2}$) the dust polarization fraction is \textit{anti-correlated} with $f_{\rm CNM}$, consistent with geometric depolarization from differently oriented magnetic fields associated with multiple CNM structures along the LOS. While the higher-density sightlines through our simulated volume qualitatively support this picture, the small volume of the simulation is not necessarily representative of the ISM path lengths probed by the higher-column density sightlines in \citet{Lei:2024}. Our main focus here is thus the interpretation of the ``diffuse" regime of \citet{Lei:2024}: and indeed, when we restrict the analysis to sightlines with fewer CNM intersections, the positive $f_{\rm CNM}$–polarization correlation emerges clearly.

Although the quantitative strength of the correlations depends on the adopted selection cuts, the qualitative trend persists across a range of $\log(N_{\rm HI})$: in regimes where $N_{\rm HI}$ and polarization fraction are themselves weakly correlated, $f_{\rm CNM}$ remains positively correlated with polarization fraction. This separation between total column effects and CNM mass fraction reinforces the interpretation that the observed trend is not simply driven by overall column density.

Having reproduced the observational trend, we next examine whether it supports the interpretation of \citet{Lei:2024} that magnetic fields are more ordered in the CNM column than in the WNM column. Our results are consistent with this picture, and allow us to probe how to map this interpretation to the three-dimensional structure of magnetic fields. 

We find that, when normalized by mass (or column density), the CNM exhibits lower magnetic disorder than the WNM. However, the characteristic spatial scale of CNM structures is much smaller than that of WNM structures. 

Thus, depending on whether disorder is measured per unit path length or per unit mass, the CNM can appear either more or less ordered. When expressed in terms of column density—the quantity most relevant for polarization weighting—the CNM contributes less angular variation per unit mass, reinforcing the conclusion that it hosts more coherent magnetic geometries despite its smaller-scale structure. Consistent with this picture, the $f_{\rm CNM}$–$p$ relation transitions from positive to anti-correlation at high column densities, where multiple independent CNM structures are more likely to be intersected along the LOS.

This distinction also has implications for the interpretation of observational measures of magnetic structure. Recent observational efforts, such as the Interstellar Polarization Structure (IPS) analyses \citep[e.g.,][]{Angarita:2024}, attempt to characterize magnetic field coherence using statistical properties of dust polarization maps. These approaches are inherently sensitive to the superposition of multiple structures along the LOS and therefore to how magnetic disorder accumulates with both path length and column density. In this context, our results suggest that part of the remaining ambiguity in interpreting such measurements may arise from the distinction between disorder per unit path length and per unit mass. Observational diagnostics that combine polarization with phase-sensitive tracers—such as \ion{H}{1} absorption/emission decomposition or velocity-resolved structure—will be essential for disentangling the relative contributions of CNM and WNM to the observed polarization signal.

Our analysis of the 3D magnetic field structure within single-phase structures requires an operational definition of a “cloud.” We identify contiguous, one-dimensional segments of a given phase along individual sightlines. This definition lies between theoretical and observational conventions. Unlike full three-dimensional structures, our clouds are defined locally along each LOS; unlike observational components identified in velocity space \citep[e.g.,][]{Panopoulou:2020, Putman:2026}, they are guaranteed to be contiguous in real space. A closer observational analog may be contiguous structures identified in three-dimensional dust maps along individual sightlines \citep[e.g.,][]{Halal:2024}, or the peak-count statistics of these maps, which show correspondence with \ion{H}{1} absorption components \citep[][]{Nowotka:2025}. While the mapping between our cloud definition and observational components is not one-to-one, the qualitative trends—particularly the relationship between phase structure, magnetic alignment, and polarization—should be robust to these definitional differences.

We also find that the simulated CNM structures are preferentially elongated along the local magnetic field. This is qualitatively consistent with observations showing that \ion{H}{1}-traced CNM is highly filamentary, preferentially cold \citep{Clark:2019a, Peek:2019, Kalberla:2025}, and aligned with the plane-of-sky magnetic field \citep{McClureGriffiths:2006, Clark:2014, Clark:2015}. In magnetized cooling flows, density enhancements, turbulent strain, and magnetic-field geometry evolve together: radiative compression and shear can produce cold atomic structures whose long axes preferentially follow the local field \citep[e.g.,][]{GrandaMunoz:2025}. In this picture, the enhanced magnetic coherence that we infer per unit mass in the CNM reflects not only its higher density, but also the anisotropic manner in which cold gas forms out of the diffuse medium. This interpretation is consistent with simulation-based studies finding systematic variations in magnetic-field strength and geometry with gas phase and density \citep[e.g.,][]{Ponnada:2022}, and is complementary to recent work on CNM filament formation in thermally unstable gas \citep[e.g.,][]{Ho:2023}. We caution, however, that our one-dimensional LOS ``cloud'' definition is operational: it should not be identified directly with either a three-dimensional filament or a velocity-decomposed observational component.

The simulations analyzed here are particularly useful for this analysis because they combine high spatial dynamic range, magnetized turbulence, and bistable radiative heating and cooling in a volume for which both the three-dimensional phase structure and the corresponding synthetic \ion{H}{1} and dust-polarization observables can be measured self-consistently \citep{Fielding:2023}. They therefore occupy a complementary regime to both specialized high-resolution studies of MHD turbulence and reconnection \citep[e.g.,][]{Beresnyak:2014,Kowal:2017} and multiphase ISM simulations that include additional Galactic physics but generally at lower effective resolution \citep[e.g.,][]{Padoan:2016,Kritsuk:2017,Kim:2017,Rathjen:2021}. For the present problem, the central advantage is that the simulations allow a controlled test of whether an $f_{\rm CNM}$--polarization correlation can arise from phase-dependent magnetic geometry rather than from total-column effects alone.

At the same time, the simulations are intentionally idealized. They describe a periodically driven turbulent box rather than a stratified Galactic environment, do not include supernova driving, gravity, detailed chemistry or ionization, explicit conduction, or explicit resistivity in the three-dimensional calculation, and they lack the large-scale magnetic and vertical structure of the Milky Way. The simulated volume is also relatively dense and turbulence-dominated, with a higher CNM fraction than is typical of high-Galactic-latitude sightlines; as a result, it likely under-samples the long WNM-dominated path lengths that contribute to real diffuse-sky observations. In addition, because the cooling model does not explicitly track ionization, the $\sim 10^4$ K gas should be interpreted as a generic warm phase rather than a clean separation between WNM and WIM. Finally, because the three-dimensional simulations rely on numerical dissipation, they should not be used to make quantitative claims about reconnection rates, current-sheet demographics, or the true dissipation scale, even though explicit 2.5D tests support the qualitative robustness of plasmoid formation \citep{Fielding:2023}. For the present analysis, the most robust conclusions are therefore qualitative: sightlines with relatively few independent CNM structures can show a positive $f_{\rm CNM}$--$p$ correlation; CNM structures are preferentially aligned with the local magnetic field; and the apparent magnetic ordering of a phase depends on whether disorder is normalized by path length or by mass/column density. The precise cloud-count threshold, column-density interval, and phase-restricted polarization distributions should be regarded as model-dependent rather than as direct Galactic predictions.
\section{Summary}
\label{sec:summary}

We used high–resolution simulations of the turbulent, magnetized, multiphase ISM to investigate the physical origin of the correlation between CNM mass fraction and dust polarization fraction reported by \citet{Lei:2024}. By generating synthetic \ion{H}{1} and dust–polarization maps, we connected observed phase–polarization correlations to the underlying three–dimensional magnetic field structure across ISM phases.
Our main conclusions are:
\begin{enumerate}

    \item \textbf{The magnetic field geometry exhibits distinct phase dependence.} The degree of magnetic disorder appears to depend specifically on gas phase, and not just on cloud geometry. While magnetic fields in the CNM are more ordered per unit mass, magnetic fields in the WNM are more ordered per unit length. Sampling the WNM material along short path lengths that are consistent with the CNM size distribution, we still find a marked difference in the degree of magnetic disorder between the CNM and WNM.

    \item \textbf{The $f_{\rm CNM}$–polarization correlation is robust.} For $19 \lesssim \log(N_{\rm HI}[\mathrm{cm}^{-2}]) \lesssim 20.5$, $f_{\rm CNM}$ is positively correlated with polarization fraction. This trend persists even when multiple CNM structures are intersected along the sightline.

    \item \textbf{CNM structures are preferentially magnetically aligned and magnetically coherent per unit mass.} CNM clouds tend to elongate along the local magnetic field. When normalized by column density or total mass, the CNM exhibits lower magnetic disorder than the UNM or WNM.

   \item \textbf{Magnetic disorder depends on the metric used.} At fixed path length, WNM segments can appear more magnetically ordered than CNM segments, reflecting the shorter characteristic scales of cold structures. At fixed mass or column density, however, the CNM exhibits lower angular variation and is the more coherent phase for polarization weighting.
\end{enumerate}

Together, these results support a picture in which diffuse CNM structures host relatively ordered magnetic fields aligned with filamentary gas, producing higher polarization fractions along CNM–dominated sightlines in the low-column density sky. 

While the qualitative trends found here agree with observations, the limited box size and driven–turbulence setup restrict quantitative accuracy, particularly for WNM column densities and large-scale magnetic coherence. Extending this analysis to stratified, galaxy-scale simulations spanning a broader range of magnetic field strengths, turbulence modes, and phase balances will be essential for testing the generality of these conclusions.

\acknowledgments
I.S.B. was supported by NASA through the Hubble Fellowship, grant HST-HF2-51525.001-A awarded by the Space Telescope Science Institute, which is operated by the Association of Universities for Research in Astronomy, Incorporated, under NASA contract NAS 5-26555.
S.E.C. acknowledges support from the National Science Foundation under grant No. AST-2441452, and support from an Alfred P. Sloan Research Fellowship. M.L. was supported by a 2MB Graduate Fellowship for Frontier Research Fund.

\bibliography{main, adsbib}

\end{document}